\documentclass[reprint,superscriptaddress,nofootinbib,amsmath,amssymb,aps]{revtex4-2}

\usepackage{graphicx}
\usepackage{dcolumn}
\usepackage{bm, amsmath, amssymb, amsthm}
\usepackage{hyperref}       
\usepackage{url}            
\usepackage{booktabs}       
\usepackage{amsfonts}       
\usepackage{nicefrac}       
\usepackage{microtype}      
\usepackage{lipsum}
\usepackage{xcolor}
\usepackage{comment}
\usepackage{subcaption}

\newcommand{\lr}[1]{\textcolor{red}{\textbf{LR}: #1}}
\newcommand{\fz}[1]{\textcolor{blue}{\textbf{FZ}: #1}}

\begin{document}

\preprint{APS/123-QED}

\title{From Scarce Functional Labels to Label-Aware Generation\\ in Homologous Protein Families
}

\author{Lorenzo Rosset}
    \affiliation{École Normale Supérieure, CNRS, Laboratoire de Physique - LPENS, Paris, France}
    \affiliation{Sorbonne Universit\'{e}, CNRS, Dept.~of Computational, Quantitative and Synthetic Biology - CQSB, Paris, France}
  
\author{Martin Weigt}
\email{martin.weigt@sorbonne-universite.fr}
    \affiliation{Sorbonne Universit\'{e}, CNRS, Dept.~of Computational, Quantitative and Synthetic Biology - CQSB, Paris, France}
    \affiliation{Institut Universitaire de France (IUF), France}

\author{Francesco Zamponi}
\email{francesco.zamponi@uniroma1.it}
    \affiliation{Dipartimento di Fisica, Sapienza Universit\`a di Roma, Rome, Italy}

\date{\today}

\begin{abstract}

Accurately annotating and controlling protein function from sequence data remains a major challenge in protein engineering, especially when functional labels are scarce within large homologous families. Here, we study a two-stage light-supervision strategy for fine-grained functional annotation and label-aware sequence generation. First, we compare several sequence representations, including one-hot encodings, Restricted Boltzmann Machines (RBMs), and ESM2-based protein language model embeddings, for predicting intra-family specificity labels from limited supervision. By using train/test splits that explicitly reduce phylogenetic leakage, we show that ESM2-based representations do not systematically outperform family-specific RBM embeddings or even simple one-hot baselines in this regime. Second, we use the inferred annotations to train an annotation-aware RBM capable of generating artificial homologs conditioned on prescribed labels. Across several protein families, we quantify how the number and quality of available labels determine the reliability of conditional generation. Our results show that scarce annotations can support label-aware protein design when they are accurately propagated, while also highlighting the importance of phylogeny-aware evaluation for assessing functional annotation methods within homologous families.
\end{abstract}

\maketitle

\section{Introduction}

The sequence-structure-function paradigm in biology posits that a protein's amino acid sequence determines its structure, which in turn underlies its function. While AlphaFold \cite{jumper2021highly} has largely resolved the sequence-to-{\it structure} problem, predicting protein {\it function} remains far more complex. In specific settings like protein-protein interactions, structure-based docking \cite{dequeker2022complete} or sequence-based machine learning methods can infer binding partners and sites \cite{cong2019protein, evans2021protein, grassmann2024protein}, including the case of interaction specificity~\cite{gueudre2016simultaneous, bitbol2016inferring, meynard2023generating}. However, when a function is less directly reflected in structure, as is often the case for enzymes, finding the sequence determinants of protein function remains a major challenge.

In the case of abundant functionally annotated data, supervised or semi-supervised learning with deep protein language model (pLMs) embeddings has proven effective to predict functions such as enzyme commission (EC) numbers \cite{doi:10.1126/science.adf2465}, catalytic activity \cite{10.1093/nar/gkae1245}, gene ontology (GO) annotations \cite{biom12111709, ProteInfer}, or protein families \cite{vitale2024evaluating}.
However, more generally, protein function lacks a clear-cut mathematical definition, and the dependence on context-specific low-throughput experiments, which are typically time and cost-intensive, makes quantitative data rather rare. In the Uniprot database \cite{uniprot2025uniprot}, out of about 150 million deposited protein sequences, less than 0.4\% contain human annotations, many of which are of structural rather than of functional nature.
Hence, precise functional annotations are, in most cases, scarce; typically, one might expect to have of the order of 100 such annotations in a protein family that might contain $10^4-10^6$ homologous sequences.

To make the most of available data, the aim of this work is to develop methods that link protein sequences (genotypes) to functional phenotypes with minimal supervision. 
We focus on a common yet challenging scenario, which manifests in two subsequent tasks:
\begin{itemize}
    \item predicting fine-grained specificities within a homologous protein family using only few labeled sequences;
    \item using these predictions for the generation of artificial homologs conditioned on a given specificity.
\end{itemize}
The ability to generate such functionally specific artificial protein sequences is of great interest for protein design, potentially in combination with subsequent refinement by directed evolution \cite{romero2009exploring}.

However, the first point already presents three major difficulties, which directly impact our ability to succeed in the second. First, the limited number of labeled sequences makes it difficult for traditional supervised methods to generalize in the high-dimensional space of homologous proteins, whose size is typically exponential in the protein length \cite{trinquier2021efficient}. Second, because homologous protein sequences share a highly similar three-dimensional structure, predicted folding offers limited additional information. 
Third, homologous protein families show complex phylogenetic correlations between sequences that, for example, make it difficult to split the dataset into meaningful test and training subsets \cite{petti2022constructing}, or to dissect functional similarity from evolutionary relatedness \cite{vicedomini2022multiple}.
In this setting, the specificities we aim to infer are subtle, fine-grained features -- unlike the broader functional categories of, e.g., GO annotations. As a result, we face a weak labeling signal, phylogenetically correlated sequences, and hardly defined functional distinctions.

In this work, we investigate the extent to which sequence embeddings can reflect the underlying functional landscape and link sequences to their specificities. A straightforward option, which goes beyond the trivial but widely used one-hot encoding, is to exploit the multiple-sequence alignment (MSA) of a protein family ($\sim 10^3$–$10^6$ sequences) to train a latent-variable model such as a Restricted Boltzmann Machine (RBM), yielding unsupervised representations that may correlate better with specificity labels \cite{tubiana2019learning, shimagaki2019selection, PhysRevE.108.014110}. 
A second and widely used option consists of exploiting the latent representations learned by a pLM encoder, pre-trained across hundreds of millions of proteins belonging to many diverse families. Intuitively, a model that has already learned the ``grammar'' of the protein language might encode general sequence features and may transfer well to narrower functional tasks, even if these are not explicitly present in the training data.
Whether or not such pre-trained pLMs can help recognize sub-functions in protein families is not a trivial question since those models are trained on the whole proteome, and specific details at the family level might be overlooked.

When a modest number of annotations is available ($\sim 10^3$ annotated sequences), one can try to improve the pLM sequence representation, aligning the embedding space with functional labels, by adding a classification head on top of the pre-trained model and adjusting the model weights until the prediction loss on a held-out validation set saturates.
We chose ESM2-650M \cite{lin2023evolutionary} as a representative pLM for its wide adoption in the literature and since its moderate number of parameters (650 million) still permits carrying out computations on a single GPU node.

We validate our approach on several protein families. The Response Regulator (RR) family (Pfam PF00072~\cite{paysan2025pfam}), involved in bacterial two-component signaling systems, serves as our main testbed, with $\sim 8 \cdot 10^5$ sequences and 8 specificity classes defined by the proteins' domain architecture. To test broader applicability, we also consider SH3 domains (PF00018, with orthology group labels \cite{lian2024deep}), the Globin family (PF00042, functional labels from UniProt \cite{ziegler2023latent}), and Chorismate Mutase (CM, PF01817) enzymes, where experimentally tested {\em in vivo} functionality in {\it E.~coli} defines the labels~\cite{kast1996exploring, russ2020evolution, netti2026expanding}.
Across these different families and labeling schemes, we show that embedding the sequences using unsupervised or self-supervised models (in particular, pLMs) does not consistently yield significant advantages in terms of label predictions when we explicitly devise the train/test splits to minimize the phylogenetic signal leakage. This finding shows that, in this fine-grained intra-family setting, the tested pLM-based representations do not automatically overcome the challenges posed by highly correlated homologous data.

As mentioned before, enriching functional labeling of protein families is not just interesting per se, but can be used to build probabilistic models -- here label-aware RBM -- for the conditional generation of artificial homologs with given labels. Besides evident biotechnological applications in protein engineering and design, such functionally resolved generative probabilistic models can be used to explore specificity-changing evolutionary trajectories \cite{frechette2021evolutionary, mauri2023transition,rehan2025design}, and to analyze the role of epistasis in protein-functional evolution \cite{di2024emergent}.
Therefore, our second major result is to show that, provided that we can obtain sufficiently high-quality labels inferred with the aforementioned strategy, generative models can identify residue motifs linked to specificities and design sequences with targeted functions. 

The structure of this paper is as follows. In Sec.~\ref{sec:pipeline}, we define the general pipeline and specify the used data embeddings and the definition of the training and test sets, which underlie the entire work.  In Sec.~\ref{sec:classification} we study the classification task. We compare the label prediction performance of a simple logistic regression model trained on the embeddings of an RBM and a pLM against the predictions obtained on the raw one-hot encoding of the family's MSA. Finally, in Sec.~\ref{sec:conditioned generation}, we imagine that we have extended the annotations to unlabeled sequences using the methods above, and we study how an RBM model responds to the accuracy and number of annotations provided during the training for the task of generating protein sequences subject to having a chosen specificity.

\begin{figure*}[t]
    \centering
    \includegraphics[width=\linewidth]{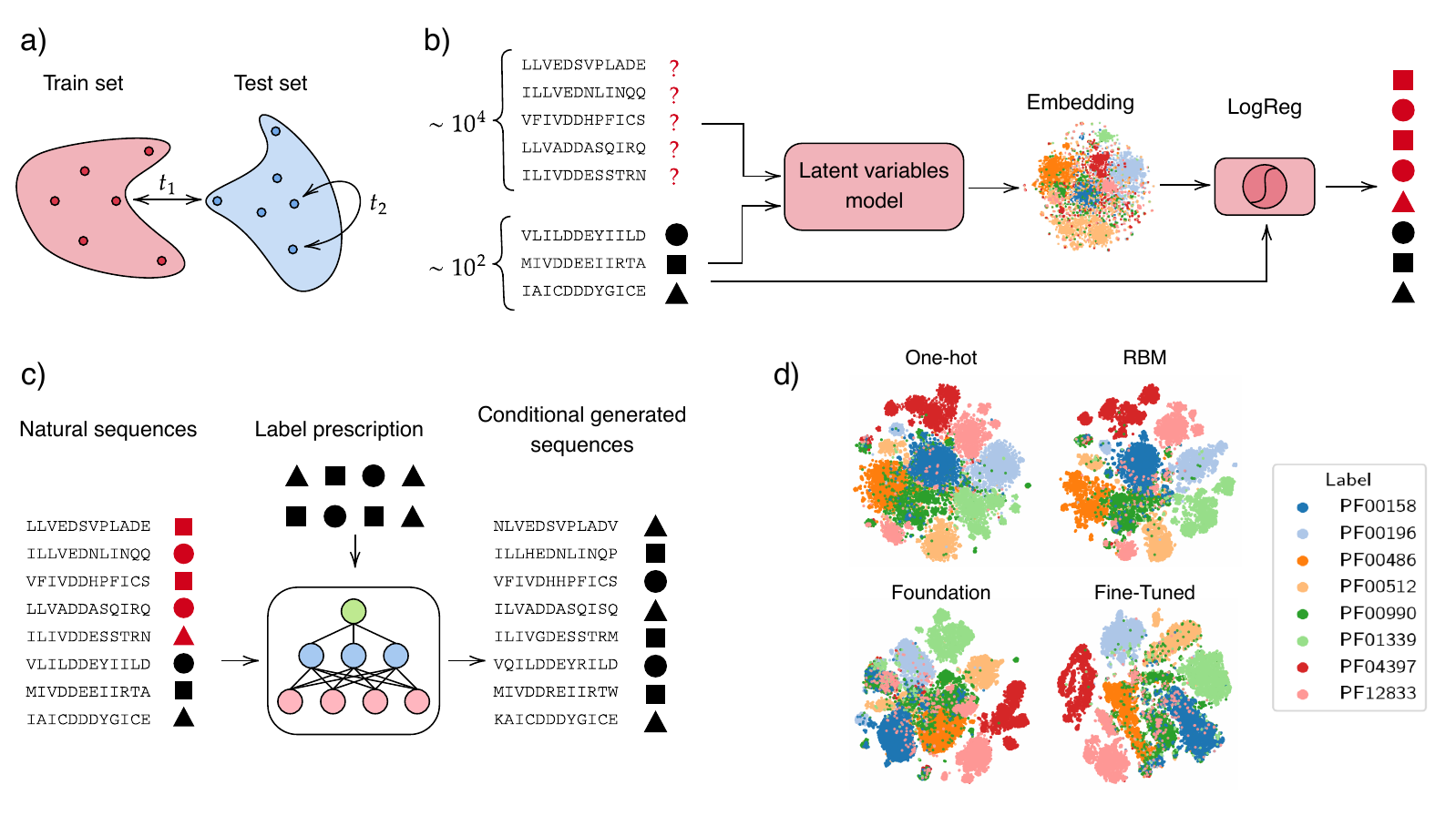}
    \caption{
    (a) Train/test splitting using \texttt{cobalt} \cite{petti2022constructing}. 
    (b) Data augmentation pipeline: sequences are embedded using a latent variables model and a logistic regression is trained on the embedding using the few labels available. The resulting predictor is used to label the remaining sequences.
    (c) Label-specific generation pipeline: natural sequences and inferred labels are used for training an RBM. After training, the model can be used to generate artificial sequences having the desired specificity. 
    (d) t-SNE clustering comparison of the test set sequence embeddings of RR. Here, the train/test splitting is done with \texttt{t1}$=0.4$, and 5000 training sequences have been used for the pLM fine-tuning.}
    \label{fig:general scheme}
\end{figure*}

\section{Results}

\subsection{Pipeline}
\label{sec:pipeline}

We consider as input data a set of homologous protein sequences that have been previously organized into either a union of full-length sequences (for pLM embedding) or an MSA (for one-hot encoding and RBM embedding). In the latter case, each sequence is represented as a string of 21 symbols (20 amino acids and one alignment gap), and all sequences have the same length. A typical data set will contain $\sim 10^4-10^6$ sequences, and we assume that a much smaller subset of $\sim 10^2 - 10^3$ sequences have been annotated for some specific functionality. Annotations are labels assigned to sequences; here we consider categorical labels that can belong to a fixed number of functionality classes, cf. {\em Methods and Materials} Sec.~\ref{sec:data} for more details on the considered families and annotation schemes. As stated in the introduction, the goals of our pipeline are to (i) annotate all the sequences in the input dataset ({\it data augmentation}), and (ii) generate new, artificial sequences that belong to each of the functionality classes ({\it label-specific generation}).

Our data augmentation and label-specific generation pipeline (Fig.~\ref{fig:general scheme}) is implemented and tested as follows.
We consider a dataset where all sequences are labeled, and we hide the labels of most sequences in the dataset to use them as a test set, using a proper train/test splitting (Fig.~\ref{fig:general scheme}a) to be discussed below. Then, we first train several `predictor' models, based on a logistic regression applied to a given embedding of the data (Fig.~\ref{fig:general scheme}b).
Second, we use the predictor to label the entire protein family and we train a `generator' model on the augmented data (Fig.~\ref{fig:general scheme}c). We use a simple RBM architecture as a benchmark for the generator.
We study how this data augmentation affects the ability of the RBM to generate sequences conditioned on a required specificity. In particular, to assess the self-consistency of the pipeline, sequences generated with a given specificity prescription are re-labeled using the predictor, and we check whether the predicted label is consistent with the one used for generation.

We compare the performance of several embedding schemes (Fig.~\ref{fig:general scheme}d) for the predictor, starting from simple one-hot encoded sequences and going to pLM embeddings:
\begin{itemize}
    \item {\tt One-hot}: In the simplest case, we use the one-hot encoding of the MSA as a dummy latent representation that serves as a baseline.
    \item {\tt RBM}: We train an embedding-RBM\footnote{This embedding-RBM should not be confused with the RBM used for generation in the second step of the pipeline.} on the entire unlabeled MSA (before train/test split) and consider as embedding the hidden representation that the RBM assigns to each test sequence.
    \item {\tt ESM2 foundation}: the zero-shot embedding obtained using the pretrained version of ESM2-650M. In this case, the full-length, unaligned sequences have been used.
    \item {\tt ESM2 finetuned}: only when the number of annotated training sequences is $\geq 1000$, we finetune ESM2-650M after putting a prediction head on top of it.
\end{itemize}
The RBM representation should be interpreted as a transductive, family-specific embedding: the model is trained on the full unlabeled MSA, including sequences that later appear in the test set without their labels. This matches the semi-supervised annotation scenario considered here, but differs from applying a model to completely unseen sequences. Note that the first three data representations do not make use of any annotation (unsupervised / zero-shot), whereas finetuning relies on label supervision (few-shot), and the embedded representation obtained with this approach will thus depend on the training set size.

Special care should be taken in performing the train/test splitting on the initial data. In fact, homologous protein families are the result of natural evolution, hence they display phylogenetic correlations, i.e.~sequences from evolutionarily related species are very similar to each other. Because of that, knowing the label of a given sequence would easily allow one to label the phylogenetically close ones, which can skew the metrics used in assessing the performances of the predictor (see the {\em Materials and Methods} Sec.~\ref{sec:cobalt} for a more detailed discussion).
We then adopt the \texttt{cobalt} train/test splitting method, specifically designed in \cite{petti2022constructing} to address this issue. The method selects test sequences that are distant from the training set and from each other (Fig.~\ref{fig:general scheme}a), by imposing a maximum allowed fractional sequence identity \texttt{t1} between training and test sequences, and a maximum allowed fractional sequence identity \texttt{t2} between test sequences. In this work, \texttt{t2} is always set to 0.7. The training sets of RR are constructed so that, whenever possible, each specificity class has the same number of samples in order to get an unbiased measure of the classification quality (SI, Fig.~\ref{fig:class abundance RR}). For the other datasets, we allowed for a moderate class imbalance when it was not possible to increase the number of training data otherwise (cf.~SI, Figs.~\ref{fig:class abundance SH3}, \ref{fig:class abundance PF00042} and \ref{fig:class abundance CM}).

\subsection{Data augmentation}
\label{sec:classification}

\begin{figure*}[ht]
    \centering
    \includegraphics[width=0.8\linewidth]{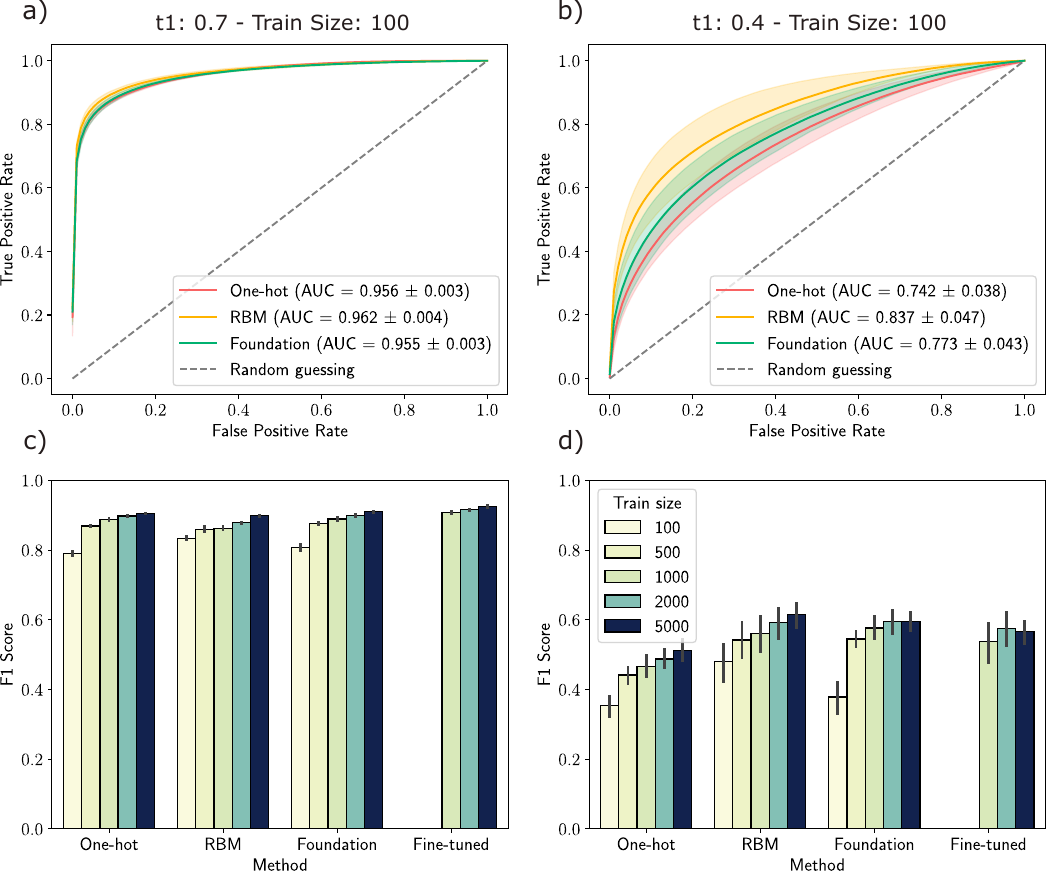}
    \caption{
    (a,b) Average ROC curves for the classification of RRs using different types of embedding, 100 training sequences and, respectively, train/test splitting thresholds of (a) \texttt{t1}=0.7 and (b) \texttt{t1}=0.4.
    (c, d) F1 scores for different methods and number of training sequences when \texttt{t1}=0.7 (c) and \texttt{t1}=0.4 (d).
    The averages have been computed over 10 independent train/test splits and the error bars/area represent one standard deviation.
    }
    \label{fig:classification PF00072}
\end{figure*}

To assess which method is better for enriching the initially limited set of annotations, we analyze the performance of the predictors with different embeddings on the classification task. We consider different numbers of training sequences and test/train splitting thresholds~\texttt{t1}. The results for the RR family are shown in Fig.~\ref{fig:classification PF00072}.  In particular, for each label, we compute the receiver operating characteristic (ROC) curve and the area under the curve (AUC) obtained by comparing the probability assigned by the classifier to each label with the ground truth. To obtain a unique measure of prediction quality, we then average the label-specific ROC curves over the different classes, and gather the statistics over 10 independent train/test \texttt{cobalt} splits. We report the label-specific ROC curves in SI, Fig.~\ref{fig:ROC labels PF00072}.

First, we consider the case where the training set has 100 labeled sequences (i.e., only slightly more than ten sequences per label), and it contains only sequences with a maximal fractional sequence identity of \texttt{t1} = 0.7. In Fig.~\ref{fig:classification PF00072}a, we show the ROC curves for different types of embeddings for the RR family. 
We report the results for the other datasets in the SI, Fig.~\ref{fig:classification other datastes}.
We see that, in general, the different embeddings have similar generalization performances. In the RR dataset, the RBM seems to be slightly better than other methods, while for the CM the pLM embedding proved to be the best representation (SI, Fig.~\ref{fig:classification other datastes}). The fact that we achieve ROC AUC of $\sim 0.95$ using only 100 training sequences regardless of the encoding strategy tells us that, for this dataset, the mere sequence similarity between sequences plays a central role in discriminating between different specificities.

The large number of homologous RR sequences allows us to use \texttt{cobalt} to produce training and test sets that are very distant from each other and yet have well-represented specificity classes. In Fig.~\ref{fig:classification PF00072}b we thus show the ROC curves of the different embeddings when the train/test splitting threshold is \texttt{t1}=0.4, again using only 100 labeled training sequences. As expected, we observe a general degradation in the prediction accuracy compared to the case with \texttt{t1}=0.7, with ROC AUC decreasing from $\sim0.96$ to $\sim0.84$ in the best case represented by the RBM embedding. Again, from this analysis, it seems that sequence encoding can provide only marginal improvements in the label predictions compared to using the simple one-hot encoding of the data. In particular, in both scenarios, the pLM embedding does not seem to capture a functional signal that generalizes beyond the RBM and the one-hot encoding of the sequences.

In Fig.~\ref{fig:classification PF00072}c and d, we show the F1 scores of each method for different numbers of training sequences. When the number of training data was $\geq 1000$, we also considered the case where a prediction head is placed on top of the pLM, and a subset of the model's weights is finetuned on the label prediction task until the loss on a left-out validation set plateaus (see SI, Sec.~\ref{sec:hyperparams} for the details). We observe a general improvement in the prediction score as we increase the number of training samples, but the gains are typically diminishing after 1000 training samples, as we will further demonstrate in the next section. When fine-tuning the pLM, we get a small improvement in the F1 score when \texttt{t1}=0.7, but the model's performance degrades when \texttt{t1}=0.4, likely due to the large mismatch between the train/validation statistics and the test set. We report in SI, tables \ref{tab:method_comparison_t1_04} and \ref{tab:method_comparison_t1_07}, the detailed scores for each model and number of training samples.

\subsection{Label-specific generation}
\label{sec:conditioned generation}

Once we have a predictor capable of annotating all sequences in the input dataset, our goal is to investigate how the accuracy of these annotations influences a generative model designed to produce artificial sequences with a desired specificity.
To this end, we consider a modification of the RBM that integrates the label information together with the aligned sequence \cite{larochelle2008classification, cossio2023disentangling, carbone2024fast}, cf.~{\em Materials and Methods} Sec.~\ref{sec: labelRBM} for details. This model can be trained on either a fully annotated dataset or a partially annotated one. After training, we can ask the model to generate artificial sequences conditioned on a chosen specificity and compare the statistical properties of the generated samples with those of the natural ones. 


\begin{figure*}[!ht]
    \centering
    \includegraphics[width=\linewidth]{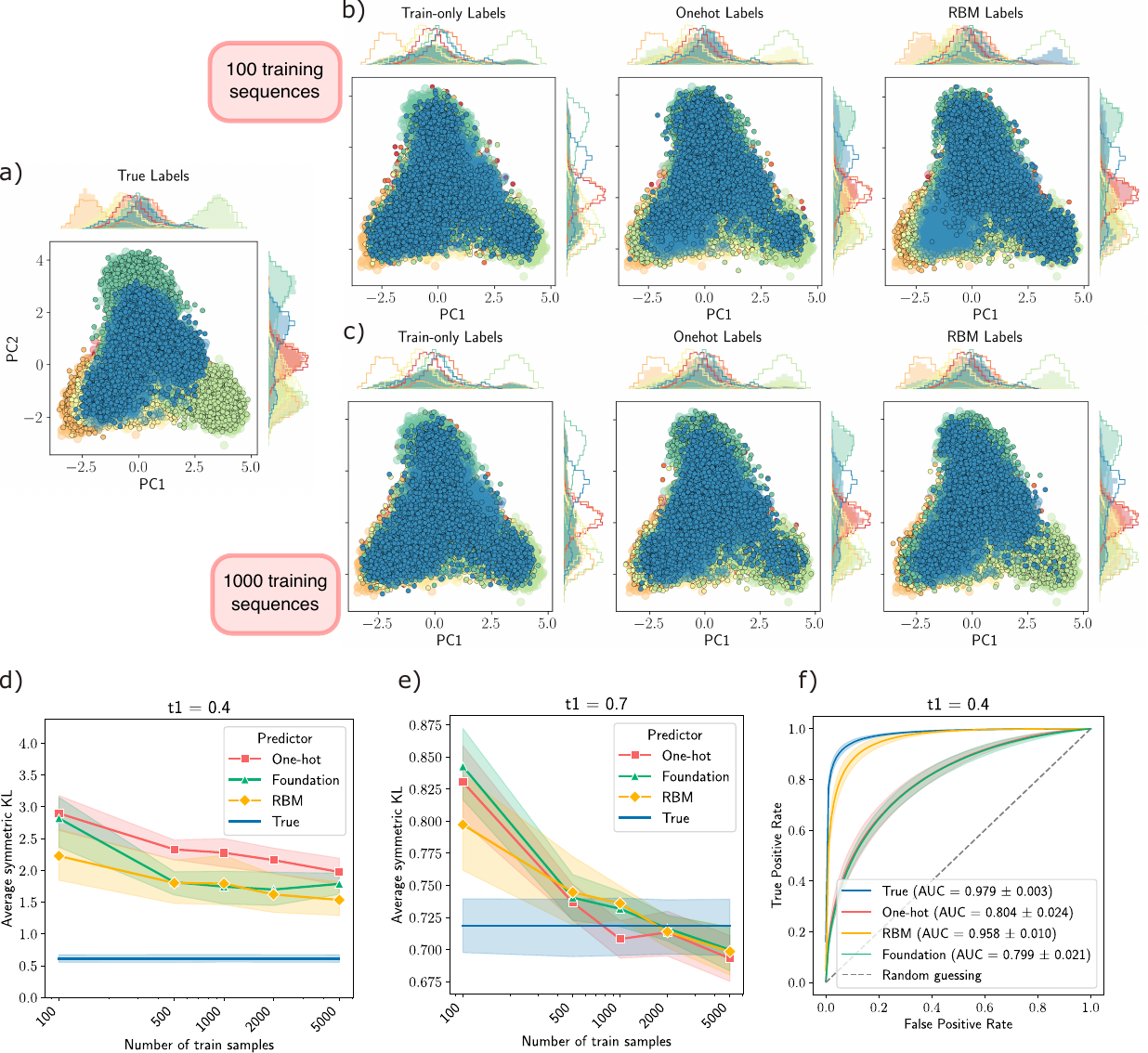}
    \caption{Conditional generation on the RR dataset. 
    (a, b, c) First two PCA projections of natural (large shaded dots and solid-line histograms) and generated (small contoured dots and shaded histograms) data for RBMs informed with different types of label. Note that all data are projected on the first two principal components of the natural MSA, to assure compatibility across datasets. The histograms of natural data, partitioned by true labels, are shown by solid lines, while the solid-filled histograms display the generated data, partitioned by the conditioning label. In (a) we show the conditional generation when all the ground-truth labels are used for the training of the RBM.
    For the predictors in the plot (b), we used \texttt{t1}=0.4 and 100 training sequences, while for the plot (c) we used \texttt{t1}=0.4 and 1000 training sequences.
    (d, e) Symmetric KL divergence between the 2D histograms of the PCA between conditionally generated samples and natural data for t1=0.4 (d) and t1 = 0.7 (e). The x-axis is reported in log-scale.
    (f) Average ROC curves quantifying the agreement between the label prescription and the label inferred on the generated data by the predictor that informed the RBM. In this case, we used \texttt{t1}=0.4 and 100 annotated training sequences.
    }
    \label{fig:generation}
\end{figure*}

We compare models trained on the union of the predictor training and test sets, while varying the amount and type of label information available during training. Specifically, we consider the following settings:
\begin{itemize}
    \item {\tt True labels} -- All sequences are annotated with their ground-truth labels. This represents the best-case benchmark, which is unrealistic in practical applications, but will provide an upper performance bound we aim at approaching with our work. 
    \item {\tt Train-only labels} -- Only the predictor training sequences are annotated with ground-truth labels, while all remaining sequences are left unannotated. This represents the worst-case baseline, which we aim to improve upon by using predicted labels.
    \item {\tt One-hot labels} -- Ground-truth labels are used for the predictor training set, whereas labels for the test set are inferred using the predictor trained on one-hot encoded sequences.
    \item {\tt RBM labels} -- As above, but test-set labels are inferred using the predictor trained on RBM embeddings.
    \item {\tt Foundation labels} -- As above, but test-set labels are inferred using predictors trained on ESM2-650M embeddings.
\end{itemize}

We first assess the quality of the generative models visually by performing a principal component analysis (PCA) of the natural and generated sequences. The principal components are computed from the input natural multiple sequence alignment (MSA), and both natural and generated sequences are projected onto this basis. Figure~\ref{fig:generation}a shows the projection onto the first two principal components for the case in which the RBM is trained using all ground-truth labels. Figures~\ref{fig:generation}b and~c show the corresponding results for the three label-inference strategies described above, using respectively 100 and 1000 annotated training sequences to train the predictors. In each panel, the natural sequences are the same and are colored according to their true labels, while the generated sequences are colored according to the label used to condition the RBM. The synthetic sequences are generated by conditioning the RBM on the same labels as those associated with the natural sequences. Marginal histograms along the two axes show the projected distributions along the first two principal components.

We observe that a fully annotated training set yields a conditional generative model that accurately reproduces the label-specific distributions of the natural data. In contrast, when only a small fraction of sequences is annotated, either 100 or 1000, the model fails to generate well-separated conditional distributions: the histograms associated with different conditioning labels largely overlap. With only 100 annotated training sequences, label inference based on the one-hot representation provides a modest improvement, but the resulting conditional generator remains inaccurate. RBM-based label predictions better reproduce the distributions for some specificities, although the overall quality of conditional generation remains limited. When the number of annotated training sequences is increased to 1000, however, our procedure yields generative models that approximate the conditional data distributions reasonably well when using either one-hot encodings or RBM embeddings (Fig.~\ref{fig:generation}c). Moreover, when a less stringent similarity threshold is used, \texttt{t1}=0.7, the conditional distributions are already well reproduced with only 100 annotated sequences (SI, Fig.~\ref{fig:conditional generation RR t1=0.7}), consistently with the strong classification performance reported in Fig.~\ref{fig:classification PF00072}a.

We then perform a quantitative analysis that also includes the Foundation-model embeddings. As shown in Fig.~\ref{fig:generation}d and~e, we compute the symmetric Kullback--Leibler (KL) divergence between the natural and generated data distributions, using two-dimensional histograms over the first two principal components. Let $P_\ell(i,j)$ and $Q_\ell(i,j)$ denote the normalized two-dimensional histograms of the natural and generated sequences, respectively, conditioned on label $\ell$, with $\ell = 1,\dots,N_\ell$, $i=1,\dots,N_x$, and $j=1,\dots,N_y$. We define
\begin{equation}
    D_{\mathrm{KL}}^{\mathrm{sym}}(P,Q)
    =
    \frac{1}{2N_\ell}
    \sum_{\ell=1}^{N_\ell}
    \left[
    D_{\mathrm{KL}}\!\left(P^\ell \,\Vert\, Q^\ell\right)
    +
    D_{\mathrm{KL}}\!\left(Q^\ell \,\Vert\, P^\ell\right)
    \right],
\end{equation}
where
\begin{equation}
    D_{\mathrm{KL}}\!\left(P^\ell \,\Vert\, Q^\ell\right)
    =
    \sum_{i=1}^{N_x}
    \sum_{j=1}^{N_y}
    P_\ell(i,j)
    \log
    \frac{P_\ell(i,j)}{Q_\ell(i,j)}.
\end{equation}
The quantity $D_{\mathrm{KL}}^{\mathrm{sym}}(P,Q)$ is non-negative and vanishes if and only if $P_\ell = Q_\ell$ for all label classes $\ell$.

This analysis confirms the qualitative observations described above. The RBM embedding is more effective than both the one-hot representation and the pLM embedding at providing label-specific information to the generative model. Comparing Fig.~\ref{fig:generation}d and~e shows that, for RR, sequence similarity plays a crucial role in enabling reliable sequence annotation and, consequently, accurate conditional generation. In particular, when the more permissive threshold \texttt{t1}=0.7 is used together with 1000 annotated sequences, the resulting conditional generation performance becomes comparable to the fully supervised setting in which all ground-truth annotations are available. Conversely, Fig.~\ref{fig:generation}d shows that for \texttt{t1}=0.4 the convergence of the different label-inference strategies toward the ground-truth benchmark is slow as the number of annotated training samples increases. This behavior can be explained by the large distance between the test and training sequences: adding more annotated training examples does not necessarily allow the model to capture the statistics of out-of-distribution test samples. Finally, the performance gains tend to plateau as the number of training labels increases, suggesting that approximately $10^3$ annotated sequences provide a favorable trade-off between the experimental cost of annotation and the reliability of conditional generation.

To evaluate the self-consistency of the proposed pipeline, we generate sequences conditioned on specific labels and then ask the corresponding predictor to relabel these synthetic sequences. Ideally, the conditioning label and the predicted label should coincide. Discrepancies can arise from two related sources of error: the predictor may misclassify a correctly generated sequence, or the generator may fail to produce a sequence consistent with the conditioning label.

To isolate the generation error, we first train a model using ground-truth labels and use it to generate approximately $3000$ sequences per specificity. We then predict the labels of these synthetic sequences using a logistic regression classifier trained on the full dataset, namely the union of the training and test sets. In Fig.~\ref{fig:generation}f, we compare the label-averaged ROC curves and AUC scores for this best-case benchmark, denoted {\tt True}, with those obtained from pipelines in which the one-hot, RBM, or Foundation predictors are used both to provide labels to the generator and to predict the specificities of the generated sequences. We omit the {\tt Train-only labels} baseline, since the previous analysis showed that it is unable to steer generation toward the desired specificities.

Under ideal conditions, the discriminator predictions are highly consistent with the target conditioning labels, yielding $\mathrm{AUC} \simeq 0.98$. RBMs informed by one-hot- and Foundation-based predictors show only partial agreement with their respective predictors, with $\mathrm{AUC} \simeq 0.80$. In contrast, self-consistency is substantially higher when labels are inferred from the RBM embedding, reaching $\mathrm{AUC} \simeq 0.96$. We hypothesize that this improvement results from the architectural alignment between the model used to infer the labels and the model used for conditional sequence generation.

\section{Discussion}

The aim of this work was twofold. First, we sought to investigate the limits of extending functional annotations to unlabeled sequences when label information is scarce and the sequences belong to the same homologous protein family. In such cases, functional distinctions are subtle and not easily captured by simple sequence similarity or overall conservation patterns. Specifically, we evaluated whether widely adopted protein language models (pLMs) provide a meaningful representation that encodes diverse biological specificities within a single family. Second, we aimed to exploit these predicted annotations to augment training data for generative probabilistic models, enabling the sampling of artificial sequences conditioned on a specific functional label.

To address the first objective, we needed to mitigate correlations arising from phylogeny and sampling biases. We constructed training sets at varying distances from the test set and homogenized the test set by removing highly similar sequences. This systematic curation allowed us to rigorously evaluate diverse protein sequence representations. We observed that embeddings derived from latent variable models—which integrate unsupervised information—generally outperform the naive MSA One-hot representation in the data annotation task, though the gains in prediction accuracy are marginal. Notably, pLM-based embeddings, which are pre-trained on broad sequence datasets without functional annotations, do not systematically capture the subtle signatures of sub-specificities within the sequences.

To better align the pLM representation with these specificities, we fine-tuned ESM2-650M by adding a prediction head and updating the final transformer layers. This procedure requires a moderate number of annotated sequences, here at least 1000, in order to reserve a validation set for early stopping. In our experiments, fine-tuning led to only modest improvements when training and test sequences were relatively close, as in the \texttt{t1} = 0.7 splits. In contrast, for the more stringent \texttt{t1} = 0.4 splits, fine-tuning degraded performance, suggesting that the validation set no longer provided a reliable proxy for generalization to distant homologs.

These results should not be interpreted as a general limitation of protein language models in functional annotation. Rather, they indicate that, in the specific regime studied here, namely fine-grained specificity prediction within homologous families under scarce supervision and phylogeny-aware splitting, the tested ESM2-based representations do not systematically outperform simpler baselines. In several cases, the family-specific RBM representation performs best. However, the RBM has access to the full unlabeled family during training, including the test sequences without their labels. This matches our semi-supervised goal, but it means that the RBM is not evaluated in the same setting as a model applied to completely unseen test sequences.

We confirmed the same qualitative picture on three additional protein families with different types of labels: orthology-group annotations for SH3 domains, functional annotations for globins, and experimentally measured binary activity for chorismate mutase. The heterogeneity of these label definitions is both a strength and a limitation of the benchmark. It shows that the conclusions are not restricted to a single dataset, but it also means that the difficulty and biological meaning of the classification task vary across families.

To address our second objective, we investigated how annotation quality affects the downstream task of conditional sequence generation. We trained label-aware RBMs under different supervision regimes: using all ground-truth labels, using only the scarce labeled subset, or using labels inferred by the different predictors. This analysis shows that conditional generation is strongly limited by the quality of the annotation step. When labels are accurate enough, the generative model can reproduce label-specific regions of sequence space and generate samples that are consistent with the prescribed specificity. Conversely, when the inferred labels are noisy or insufficient, the conditioning signal deteriorates and the generated distributions no longer match the corresponding natural label classes.

In the most challenging RR setting, with \texttt{t1} = 0.4 and only 100 labeled sequences, none of the augmentation strategies was sufficient to obtain reliable conditional generation across all classes. Increasing the number of labeled sequences improved performance, with substantial gains around 1000 labels and diminishing returns beyond that point. Among the tested representations, RBM-based label inference provided the most consistent downstream generation, possibly because the predictor and generator share a closely related latent-variable structure.

Overall, these results provide a controlled assessment of how scarce functional annotations can be propagated and used for label-aware protein generation. By explicitly accounting for phylogenetic correlations, we show that fine-grained intra-family specificity prediction remains demanding, and that high-capacity pretrained representations do not automatically outperform family-adapted or simple sequence-based baselines. At the same time, when inferred annotations are sufficiently reliable, they can effectively support conditional generative modeling. This establishes a practical light-supervision framework for connecting sparse functional measurements to targeted exploration of protein sequence space, while making explicit the annotation-quality requirements on which such strategies depend.

\section{Materials and Methods}

\subsection{Datasets: protein families and functional annotations}
\label{sec:data}

\subsubsection{Response Regulators}
The Response Regulator Receiver Domain (RR, Pfam accession: PF00072) is among the largest protein families, encompassing over 2 million sequences. Most RR are part of a two-component signal transduction system, where they are activated by a sensor histidine kinase following an environmental stimulus, subsequently triggering a transcriptional response as transcription factors. Specific functions of proteins containing RR domains can be deduced from their Pfam domain architecture, in particular via the presence of different families of DNA binding, sigma factor or histidine kinase domains. Here we present only the RR domain, which is common to all these proteins, labeled by the other functional domains. For our study, we selected 855{,}395 RR domain sequences from 8 abundant domain architectures and aligned them into a multiple sequence alignment (MSA) of length $L=112$ residues. The abundance of annotated sequences allowed us to generate training and test sets with a maximum of 40\% sequence identity, and training set sizes of 100, 500, 1{,}000, 2{,}000, and 5{,}000 sequences, each balanced across label classes whenever possible.

\subsubsection{SH3 domain}
Src-homology 3 domains (SH3, Pfam accession: PF00018) \cite{musacchio1992crystal} are small protein interaction modules commonly found in a wide range of intracellular and membrane-associated proteins. They play key roles in various cellular processes, including promoting local protein concentration, altering subcellular localization, and facilitating the assembly of large multiprotein complexes. Following the dataset from \cite{lian2024deep}, we compiled 7{,}861 SH3 domain sequences along with their corresponding multiple sequence alignment (MSA) of length $L=62$. Each sequence is annotated by orthology group, spanning 15 distinct classes. To evaluate model performance, we created training and test sets with sequence similarities capped at 70\%, and various training sizes. The test set size for various splits lies in the range of 150 - 180 sequences, balanced across 15 classes. The class distribution for one particular split is shown in SI, Fig.~\ref{fig:class abundance SH3}.

\subsubsection{Globin family}
Globins (Pfam accession: PF00042) are heme-containing proteins that function primarily in oxygen binding and transport. We trained the embedding-RBM on the MSA of the full globin family obtained from Pfam, which includes 29{,}211 sequences of length $L = 117$. Functional annotations were taken from~\cite{ziegler2023latent}, providing a total of 4{,}240 labeled sequences. Using \texttt{cobalt} with a threshold \texttt{t1}$= 0.7$, we generated train/test splits for various train set sizes and around 160 test sequences across 7 specificity classes (see SI, Fig.~\ref{fig:class abundance PF00042}).

\subsubsection{Chorismate Mutase}
Chorismate Mutases (CM, Pfam accession: PF01817)~\cite{kast1996exploring} are enzymes that catalyze the conversion of chorismate to prephenate, a key step in the biosynthetic pathway for phenylalanine and tyrosine. We used functional data from~\cite{russ2020evolution} and~\cite{netti2026expanding}, in which the authors compile a set of, respectively, 1{,}130 and 4{,}892 sequences that were experimentally tested for activity. Each sequence is annotated with a binary label indicating its \textit{in vivo} functionality in \textit{E.~coli}. From this dataset, we constructed several training and test sets with a maximum of 70\% pairwise sequence identity. The distribution of functional labels is shown in SI, Fig.~\ref{fig:class abundance CM}.

\subsection{Train/test splitting \\of homologous protein families}
\label{sec:cobalt}

Working with homologous protein families poses the problem of finding a sensible splitting of the data in a training and a test set. Because phylogenetic relationships introduce correlations into the data, 
any naive random splitting of the dataset yields test sequences which are phylogenetically close to training sequences. Consequently, the model scores similarly on the training and the test set, thereby invalidating any estimate of the model's generalization capabilities. On top of this, homologous sequences tend to be distributed unevenly across sequence space, resulting in clusters with high sequence identity. This is problematic when we want to assess the model on a test set because some regions of sequence space are overrepresented, 
preventing us from getting a global, unbiased measure of the model's performance.

To address this problem, we use the \texttt{cobalt} algorithm introduced in \cite{petti2022constructing}. The method consists of two stages. In the first stage, a training and a test set are extracted from the initial database so that no sequence in the test set has more than \texttt{t1} fractional sequence identity with any other sequence in the training set. This allows us to reduce correlations between the two sets of sequences. In the second stage, the test set is pruned until the fractional pairwise sequence identity between any pair of test sequences does not exceed \texttt{t2}. This second part of the algorithm allows us to obtain a homogeneous test set that is not skewed towards any sequence. In our experiments, we considered \texttt{t1} $\in \{0.4, 0.7\}$ and \texttt{t2}$=0.7$.

More precisely, we perform a \texttt{cobalt} splitting independently for each annotation class.
For each dataset that we tested, we followed the following train/test splitting pipeline.
\begin{itemize}
    \item We split the annotated sequences into several groups, one for each annotation class.
    \item For each group, we obtain a train/test splitting using \texttt{cobalt} with specific values for \texttt{t1} and \texttt{t2}.
    \item We merge the obtained train and test splits.
    \item We remove the label classes that do not contain at least 10 samples, both in the training set and in the test set. Then, we ensure that each class of the test set has an equal number of sequences and that each class of the training set has at most a number of sequences equal to the total number of available training samples divided by the number of classes.
\end{itemize}
Note that by first partitioning the sequences into sets of common labeling and then applying \texttt{cobalt} to each label, we are not guaranteed that no test sequence is closer than \texttt{t1} to the training set in the final split. In fact, our procedure allows sequences with {\em different} labels to be closer than \texttt{t1}. This is, however, a stricter way of testing the pipeline, because a model might mistakenly infer the same label for sequences of different specificities, because of being fooled by their sequence similarity.

\subsection{Embedding Restricted Boltzmann Machine}

The RBM is the simplest generative model implementing latent variables \cite{smolensky1986information, hinton2002training}. In the RBM, the physical (also called \textit{visible}) variables $\pmb a$ are connected to the latent (or \textit{hidden}) variables $\pmb x$ on a bipartite graph. When dealing with biological sequences, visible units are vectors of categorical (or Potts) variables representing the aligned sequences $\pmb a = \{a_1, \dots, a_L\}$ with $a_i \in\{1, \dots,q\}$ where $L$ is the size of the alignment and $q=21$ is the total number of symbols (20 amino acids plus the alignment gap). In our implementation, the latent variables are binary vectors $\pmb x \in \{0,1\}^P$ with $P$ being the chosen number of hidden units. The joint probability distribution of visible and hidden variables is a Boltzmann measure
\begin{equation}
    p(\pmb{a}, \pmb{x}) = \frac{1}{Z} e^{-\mathcal{H}(\pmb{a}, \pmb{x})} \ ,
\end{equation}
where $Z$ is a normalization constant and $\mathcal{H}$ is the statistical energy given by
\begin{equation}\label{eq:HRBMemb}
\begin{split}
    \mathcal{H}(\pmb a, \pmb x) =& - \sum_{i=1}^L \theta_i(a_i) -\sum_{\mu=1}^{P} \eta_\mu x_\mu \\ & -\sum_{i=1}^L \sum_{\mu=1}^{P} W_{i \mu}(a_i) x_\mu \ .
\end{split}\end{equation}
In this expression, $\pmb \theta$ and $\pmb \eta$ are local biases, while $W$ is a rank-3 tensor that couples visible and latent variables. The role of visible biases is to model the conservation signal in the MSA, while the hidden bias and the coupling matrix undirectly encode the higher-order epistatic signal between residues \cite{decelle2025inferring}.
The model parameters are tuned to maximize the likelihood of the data, which can be achieved via stochastic gradient ascent. The log-likelihood gradient is estimated using Markov Chain Monte Carlo (MCMC) simulations, and typically this is done using the Persistent Contrastive Divergence (PCD) scheme \cite{tieleman2008training, decelle2021equilibrium}. Here, in order to obtain models that are trained as close as possible to the equilibrium, we implement the more advanced scheme called PTT \cite{ICLR2025_4fe18591, bereux2025training}, where configurations are exchanged along the training trajectory similarly to what happens in parallel tempering.
Importantly, the bipartite structure of the RBM interaction graph allows us to sample all at once, alternatively, the visible and the hidden variables from the conditional probability distributions. Such a sampling scheme is called \textit{block Gibbs sampling}.

For each dataset, we trained the RBM for $10^5$ updates using an adaptive learning rate, 2000 Markov chains and a latent dimension of 1024.
After training, we encode train and test sequences through the deterministic map $m_\mu = p( x_\mu| \pmb a)$, forming the $P=1024$-dimensional vector $\pmb m = (m_1,...,m_P)$ representing the latent probability profile of the sequence.
We then feed this vector to a logistic regression model to predict the label.
As we show in SI (Fig.~\ref{fig:rbm scores vs time}), the predictive performance of the model increases during the training, then it drops a bit in correspondence with a major phase transition of the model and finally it grows again up to reaching a plateau. For all our predictions, we made sure to be well inside the saturating region.

\subsection{Embedding using protein language models}
Despite being appealing for its simplicity, the shallowness of the RBM architecture poses a limitation on the model representation capabilities. 
Indeed, it is known that many hidden layers are needed in order to encode high-level hierarchical features of the data. 
We thus explore the possibility of leveraging the representation capability of deep pre-trained protein language models to discriminate different specificities in homologous sequences. 
We consider ESM2-650M~\cite{lin2022language}, which consists of an encoder-only Transformer architecture trained using masked language modeling on more than 65 million unique sequences. Despite being trained on a completely different task than recognizing protein sequence specificity, the pre-trained model is capable of extracting high-level features from the data to produce a meaningful internal representation that can be used to inform downstream tasks like supervised label prediction. Given an input sequence, we consider its representative embedding as the average over the sequence positions of the model hidden representation (\textit{mean-pooling}) at the last hidden layer. In this way, every sequence is mapped in a 1280-dimensional vector that we then feed to a logistic regression model. In this work, we denote as \textit{foundation} (also called \textit{zero-shot} in the literature) the embedding produced by the pre-trained pLM.

\subsection{Finetuning the pLM on a Classification Task}\label{sec:contrastive fine-tuning}
To investigate whether fine-tuning the ESM2-650M model could better align its latent space with the distinct specificities of the RR dataset, we evaluated training set sizes of 1000, 2000, and 5000 sequences. For each condition, we randomly reserved 20\% of the sequences to serve as a validation set. We then appended a classification head—comprising a simple linear layer followed by a softmax activation—to the model's final hidden layer. Keeping all but the last six layers of ESM2-650M frozen, we trained the network to predict the ground-truth annotations of the training data. Training proceeded until the validation loss plateaued, utilizing an early stopping patience of 3 epochs. Because our primary objective was a qualitative assessment of the pLM's fine-tuning potential, we bypassed an exhaustive hyperparameter search in favor of a standard, reasonable set of defaults. Full details regarding our hyperparameter choices are reported in SI, Table~\ref{tab:hyperparams}.

\subsection{Generative Restricted Boltzmann Machine}
\label{sec: labelRBM}

\begin{figure}[t]
    \centering
    \includegraphics[width=0.7\linewidth]{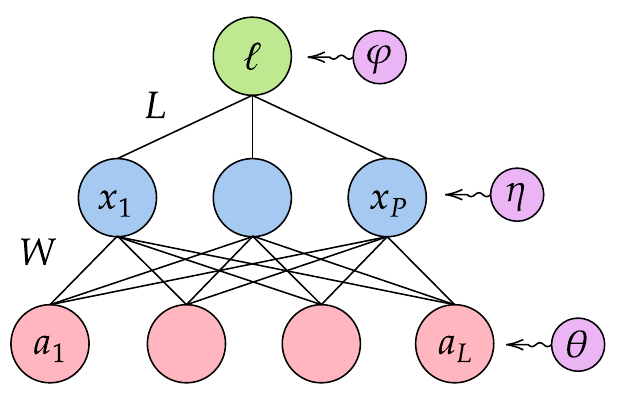}
    \caption{Schematic representation of the 
    label-specific generative RBM.
    }
    \label{fig:labelRBM}
\end{figure}

With a simple modification in the RBM architecture, it is possible to integrate the information on the label along with the amino acid sequence \cite{larochelle2008classification, carbone2024fast, AGLIARI2024106389, Alemanno_2023}.
We introduce the label variable $\ell \in \{1, \dots, K\}$ with $K$ being the number of annotation classes. The energy function of the model gains an additional term with respect to Eq.~\eqref{eq:HRBMemb},
\begin{equation}
    \mathcal{H}^A(\pmb a, \pmb x, \ell) = \mathcal{H}(\pmb a, \pmb x) + \mathcal{H}^L(\pmb x, \ell) \ ,
\end{equation}
where
\begin{equation}
    \mathcal{H}^L(\pmb x, \ell) = - \varphi(\ell)-\sum_{\mu=1}^P L_{\mu}(\ell) x_\mu \ .
\end{equation}
Here, $\pmb \varphi$ are the biases acting on the labels and $L$ is a weight matrix that couples each label with the hidden layer. In Fig.~\ref{fig:labelRBM} we report a schematic representation of the model's architecture.

The annotation-aided RBM can be trained both on fully or partially-annotated datasets. In the latter case, it is sufficient to multiply the label bias and coupling matrix by zeros to account for missing annotations. Once the model has been trained, it is possible to generate new sequences conditioned on a chosen specificity. To achieve this, we clamp the label variable to a given value and we run a Monte Carlo simulation using block Gibbs sampling between hidden and visible layers. Fixing the label has the effect of biasing the latent vector toward a label-specific pattern, which in turn constrains the simulation in the desired region of the protein sequence space.

For all the datasets, we trained the model for $5\cdot 10^4$ gradient updates using PCD with 500 hidden units, learning rate of $5 \cdot 10^{-3}$, 10 block Gibbs sampling steps between updates and 5000 Markov chains running in parallel.

\section{Data availability}
The datasets used in this work are publicly available through a \href{https://doi.org/10.5281/zenodo.20719564}{Zenodo repository}.

\section{Code availability}
The code that allows to reproduce the results of this paper is available at the following \href{https://github.com/spqb/homolog-label-generation.git}{Github repository}.

\section{Acknowledgements}

We thank Aur\'elien Decelle and Beatriz Seoane for very important discussions that helped us design the project. We thank Alessandra Carbone for useful discussions on the pLM fine-tuning.
We also thank Cl\'ement Nizak, Olivier Rivoire and Shoichi Yip for useful discussions about experimental settings for functional annotations and for feedback on our work.
We acknowledge financial support from the Horizon Europe MSCA Staff Exchange project  ``SIMBAD'' (grant agreement no. 101131463).

\bibliography{references}

\newcommand{\beginsupplement}{
        \clearpage
        \setcounter{section}{0}
        \renewcommand{\thesection}{S\arabic{section}}
        \setcounter{equation}{0}
        \renewcommand{\theequation}{S\arabic{equation}}
        \setcounter{table}{0}
        \renewcommand{\thetable}{S\arabic{table}}
        \setcounter{figure}{0}
        \renewcommand{\thefigure}{S\arabic{figure}}
     }
\beginsupplement

\onecolumngrid

\centerline{\large\bf SUPPLEMENTARY INFORMATION}
\bigskip

\twocolumngrid

\section{Hyperparameters for pLM fine-tuning}\label{sec:hyperparams}
To fine-tune the ESM2-650M model, we partitioned the data into an 80/20 training and validation split. A linear classification head, dimensioned as $\textit{hidden size} \times \textit{num. classes}$, was appended to the final hidden layer and followed by a softmax activation. We updated only the final six Transformer layers, optimizing the categorical cross-entropy loss via the AdamW optimizer with a linearly decaying learning rate. Training proceeded until the validation loss plateaued, governed by an early stopping patience of three epochs. The complete set of hyperparameters is detailed in Tab.~\ref{tab:hyperparams}.

\begin{table}[ht]
    \centering
    \begin{tabular}{cc}
    \toprule
    \textbf{Hyperparameter} & \textbf{Value} \\
    \midrule
     backbone & \texttt{facebook/esm2\_t33\_650M\_UR50D} \\
     batch size & 16 \\
     max epochs & 50 \\
     max length & 256 \\
     learning rate & $2 \cdot 10^{-5}$ \\
     weight decay & $0.001$ \\
     patience & 3 \\
     Dropout (head) & 0.1 \\
     \bottomrule
    \end{tabular}
    \caption{Hyperparameters used for the supervised fine-tuning.}
    \label{tab:hyperparams}
\end{table}

\section{Results for other families}\label{sec:other families}
\begin{figure*}[ht]
    \centering
    \includegraphics[width=\linewidth]{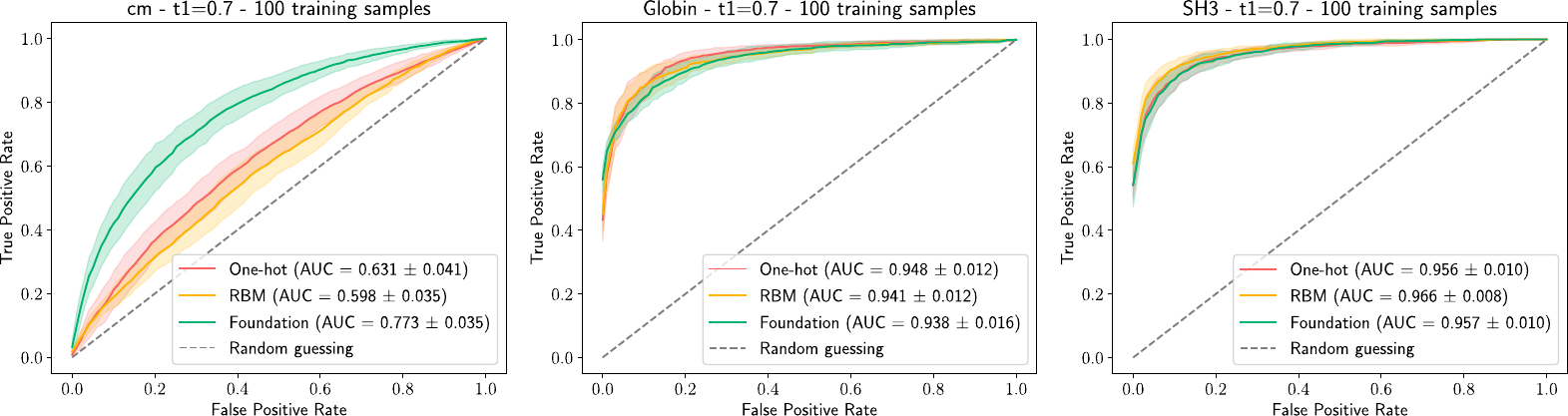}
    \caption{Mean and standard deviation of ROC curves and AUC for a) Chorismate Mutase, b) Globin family and c) Sh3 over 10 independent train/test splits. The train/test threshold is set to \texttt{t1}=0.7 and the number of training sequences is 100.}
    \label{fig:classification other datastes}
\end{figure*}

The results discussed in the main text are especially true when dealing with a training set that is at a long distance from the training sequences (RR, \texttt{t1}=0.4). In Fig.~\ref{fig:classification other datastes} we show that our findings hold true for other protein families with very different types of annotations (see Main Text, Sec.~\ref{sec:data}) even when \texttt{t1}=0.7.
Detailed results are given in Tables~\ref{tab:method_comparison_t1_04}-\ref{tab:method_comparison_sh3_t1_07}.

\begin{table*}[t]
\centering
\tiny
\setlength{\tabcolsep}{2.5pt}
\renewcommand{\arraystretch}{0.85}

\begin{subtable}[t]{0.49\textwidth}
\centering
\caption{RR, \texttt{t1}=0.4}
\label{tab:method_comparison_t1_04}
\resizebox{\linewidth}{!}{%
\begin{tabular}{llcc}
\toprule
\textbf{Train} & \textbf{Method} & \textbf{F1} & \textbf{ROC AUC} \\
\midrule
100 & Foundation & $0.378 \pm 0.075$ & $0.778 \pm 0.045$ \\
    & One-hot    & $0.355 \pm 0.051$ & $0.748 \pm 0.042$ \\
    & RBM        & $\mathbf{0.481} \pm 0.092$ & $\mathbf{0.839} \pm 0.047$ \\
\midrule
500 & Foundation & $\mathbf{0.544} \pm 0.038$ & $0.873 \pm 0.023$ \\
    & One-hot    & $0.441 \pm 0.038$ & $0.806 \pm 0.024$ \\
    & RBM        & $0.542 \pm 0.086$ & $\mathbf{0.875} \pm 0.035$ \\
\midrule
1000 & Fine-tuned & $0.538 \pm 0.097$ & $0.837 \pm 0.039$ \\
     & Foundation & $\mathbf{0.576} \pm 0.056$ & $\mathbf{0.887} \pm 0.025$ \\
     & One-hot    & $0.467 \pm 0.049$ & $0.829 \pm 0.027$ \\
     & RBM        & $0.560 \pm 0.088$ & $\mathbf{0.887} \pm 0.033$ \\
\midrule
2000 & Fine-tuned & $0.575 \pm 0.080$ & $0.868 \pm 0.030$ \\
     & Foundation & $\mathbf{0.594} \pm 0.057$ & $0.895 \pm 0.020$ \\
     & One-hot    & $0.487 \pm 0.048$ & $0.845 \pm 0.023$ \\
     & RBM        & $0.593 \pm 0.074$ & $\mathbf{0.899} \pm 0.021$ \\
\midrule
5000 & Fine-tuned & $0.565 \pm 0.053$ & $0.851 \pm 0.022$ \\
     & Foundation & $0.594 \pm 0.044$ & $0.897 \pm 0.018$ \\
     & One-hot    & $0.512 \pm 0.049$ & $0.859 \pm 0.023$ \\
     & RBM        & $\mathbf{0.614} \pm 0.062$ & $\mathbf{0.905} \pm 0.018$ \\
\bottomrule
\end{tabular}}
\end{subtable}
\hfill
\begin{subtable}[t]{0.49\textwidth}
\centering
\caption{RR, \texttt{t1}=0.7}
\label{tab:method_comparison_t1_07}
\resizebox{\linewidth}{!}{%
\begin{tabular}{llcc}
\toprule
\textbf{Train} & \textbf{Method} & \textbf{F1} & \textbf{ROC AUC} \\
\midrule
100 & Foundation & $0.807 \pm 0.018$ & $0.956 \pm 0.004$ \\
    & One-hot    & $0.790 \pm 0.009$ & $0.957 \pm 0.003$ \\
    & RBM        & $\mathbf{0.834} \pm 0.010$ & $\mathbf{0.962} \pm 0.004$ \\
\midrule
500 & Foundation & $\mathbf{0.876} \pm 0.005$ & $\mathbf{0.979} \pm 0.002$ \\
    & One-hot    & $0.869 \pm 0.004$ & $\mathbf{0.979} \pm 0.002$ \\
    & RBM        & $0.859 \pm 0.012$ & $0.975 \pm 0.004$ \\
\midrule
1000 & Fine-tuned & $\mathbf{0.907} \pm 0.007$ & $0.981 \pm 0.002$ \\
     & Foundation & $0.889 \pm 0.006$ & $\mathbf{0.983} \pm 0.001$ \\
     & One-hot    & $0.888 \pm 0.004$ & $0.983 \pm 0.001$ \\
     & RBM        & $0.863 \pm 0.007$ & $0.979 \pm 0.001$ \\
\midrule
2000 & Fine-tuned & $\mathbf{0.916} \pm 0.003$ & $0.984 \pm 0.002$ \\
     & Foundation & $0.899 \pm 0.005$ & $\mathbf{0.986} \pm 0.001$ \\
     & One-hot    & $0.898 \pm 0.004$ & $0.985 \pm 0.001$ \\
     & RBM        & $0.878 \pm 0.004$ & $0.983 \pm 0.001$ \\
\midrule
5000 & Fine-tuned & $\mathbf{0.925} \pm 0.003$ & $0.987 \pm 0.002$ \\
     & Foundation & $0.910 \pm 0.002$ & $\mathbf{0.989} \pm 0.001$ \\
     & One-hot    & $0.904 \pm 0.002$ & $0.986 \pm 0.000$ \\
     & RBM        & $0.898 \pm 0.003$ & $0.988 \pm 0.001$ \\
\bottomrule
\end{tabular}}
\end{subtable}

\caption{Classification performance for the RR dataset at two train/test identity thresholds. Mean and standard deviation are reported over independent train/test splits; the best method for each setting is highlighted in bold.}
\label{tab:method_comparison_rr}
\end{table*}

\begin{table*}[t]
\centering
\tiny
\setlength{\tabcolsep}{2pt}
\renewcommand{\arraystretch}{0.82}

\begin{subtable}[t]{0.32\textwidth}
\centering
\caption{CM, \texttt{t1}=0.7}
\label{tab:method_comparison_cm_t1_07}
\resizebox{\linewidth}{!}{%
\begin{tabular}{llcc}
\toprule
\textbf{Train} & \textbf{Method} & \textbf{F1} & \textbf{AUC} \\
\midrule
100 & Foundation & $\mathbf{0.705} \pm 0.029$ & $\mathbf{0.773} \pm 0.037$ \\
    & One-hot & $0.597 \pm 0.029$ & $0.631 \pm 0.043$ \\
    & RBM & $0.569 \pm 0.033$ & $0.598 \pm 0.037$ \\
\midrule
500 & Foundation & $\mathbf{0.719} \pm 0.023$ & $\mathbf{0.809} \pm 0.016$ \\
    & One-hot & $0.589 \pm 0.033$ & $0.639 \pm 0.032$ \\
    & RBM & $0.564 \pm 0.028$ & $0.620 \pm 0.030$ \\
\midrule
1000 & Foundation & $\mathbf{0.717} \pm 0.020$ & $\mathbf{0.811} \pm 0.018$ \\
     & One-hot & $0.566 \pm 0.027$ & $0.621 \pm 0.033$ \\
     & RBM & $0.552 \pm 0.025$ & $0.624 \pm 0.026$ \\
\midrule
2000 & Foundation & $\mathbf{0.701} \pm 0.024$ & $\mathbf{0.818} \pm 0.019$ \\
     & One-hot & $0.540 \pm 0.024$ & $0.609 \pm 0.024$ \\
     & RBM & $0.543 \pm 0.023$ & $0.626 \pm 0.025$ \\
\midrule
5000 & Foundation & $\mathbf{0.674} \pm 0.020$ & $\mathbf{0.823} \pm 0.021$ \\
     & One-hot & $0.505 \pm 0.024$ & $0.606 \pm 0.025$ \\
     & RBM & $0.500 \pm 0.016$ & $0.621 \pm 0.020$ \\
\bottomrule
\end{tabular}}
\end{subtable}
\hfill
\begin{subtable}[t]{0.32\textwidth}
\centering
\caption{Globin, \texttt{t1}=0.7}
\label{tab:method_comparison_globin_t1_07}
\resizebox{\linewidth}{!}{%
\begin{tabular}{llcc}
\toprule
\textbf{Train} & \textbf{Method} & \textbf{F1} & \textbf{AUC} \\
\midrule
100 & Foundation & $\mathbf{0.772} \pm 0.021$ & $0.938 \pm 0.017$ \\
    & One-hot & $0.761 \pm 0.042$ & $\mathbf{0.948} \pm 0.013$ \\
    & RBM & $0.766 \pm 0.056$ & $0.941 \pm 0.013$ \\
\midrule
500 & Foundation & $\mathbf{0.807} \pm 0.028$ & $0.948 \pm 0.012$ \\
    & One-hot & $0.787 \pm 0.033$ & $0.949 \pm 0.010$ \\
    & RBM & $0.783 \pm 0.041$ & $\mathbf{0.961} \pm 0.012$ \\
\midrule
1000 & Foundation & $\mathbf{0.812} \pm 0.020$ & $0.955 \pm 0.012$ \\
     & One-hot & $0.785 \pm 0.023$ & $0.949 \pm 0.010$ \\
     & RBM & $0.782 \pm 0.041$ & $\mathbf{0.963} \pm 0.009$ \\
\midrule
2000 & Foundation & $\mathbf{0.804} \pm 0.025$ & $0.956 \pm 0.013$ \\
     & One-hot & $0.779 \pm 0.022$ & $0.947 \pm 0.009$ \\
     & RBM & $0.767 \pm 0.033$ & $\mathbf{0.963} \pm 0.010$ \\
\midrule
5000 & Foundation & $\mathbf{0.802} \pm 0.033$ & $0.956 \pm 0.014$ \\
     & One-hot & $0.761 \pm 0.025$ & $0.944 \pm 0.010$ \\
     & RBM & $0.764 \pm 0.036$ & $\mathbf{0.962} \pm 0.008$ \\
\bottomrule
\end{tabular}}
\end{subtable}
\hfill
\begin{subtable}[t]{0.32\textwidth}
\centering
\caption{SH3, \texttt{t1}=0.7}
\label{tab:method_comparison_sh3_t1_07}
\resizebox{\linewidth}{!}{%
\begin{tabular}{llcc}
\toprule
\textbf{Train} & \textbf{Method} & \textbf{F1} & \textbf{AUC} \\
\midrule
100 & Foundation & $0.684 \pm 0.057$ & $0.957 \pm 0.011$ \\
    & One-hot & $0.704 \pm 0.044$ & $0.956 \pm 0.010$ \\
    & RBM & $\mathbf{0.760} \pm 0.040$ & $\mathbf{0.966} \pm 0.008$ \\
\midrule
500 & Foundation & $0.805 \pm 0.029$ & $0.975 \pm 0.006$ \\
    & One-hot & $0.806 \pm 0.024$ & $0.972 \pm 0.007$ \\
    & RBM & $\mathbf{0.836} \pm 0.010$ & $\mathbf{0.979} \pm 0.005$ \\
\midrule
1000 & Foundation & $0.834 \pm 0.036$ & $0.979 \pm 0.004$ \\
     & One-hot & $0.814 \pm 0.037$ & $0.974 \pm 0.006$ \\
     & RBM & $\mathbf{0.849} \pm 0.016$ & $\mathbf{0.981} \pm 0.004$ \\
\midrule
2000 & Foundation & $0.838 \pm 0.033$ & $0.980 \pm 0.004$ \\
     & One-hot & $0.825 \pm 0.038$ & $0.975 \pm 0.008$ \\
     & RBM & $\mathbf{0.845} \pm 0.022$ & $\mathbf{0.981} \pm 0.005$ \\
\midrule
5000 & Foundation & $0.833 \pm 0.032$ & $0.980 \pm 0.004$ \\
     & One-hot & $0.821 \pm 0.040$ & $0.974 \pm 0.008$ \\
     & RBM & $\mathbf{0.847} \pm 0.017$ & $\mathbf{0.981} \pm 0.006$ \\
\bottomrule
\end{tabular}}
\end{subtable}

\caption{Classification performance for CM, Globin, and SH3 at \texttt{t1}=0.7. Mean and standard deviation are reported over independent train/test splits; the best method for each setting is highlighted in bold.}
\label{tab:method_comparison_other_families}
\end{table*}

\section{Additional figures}

\begin{figure*}[t]
    \centering
    \includegraphics[width=0.8\linewidth]{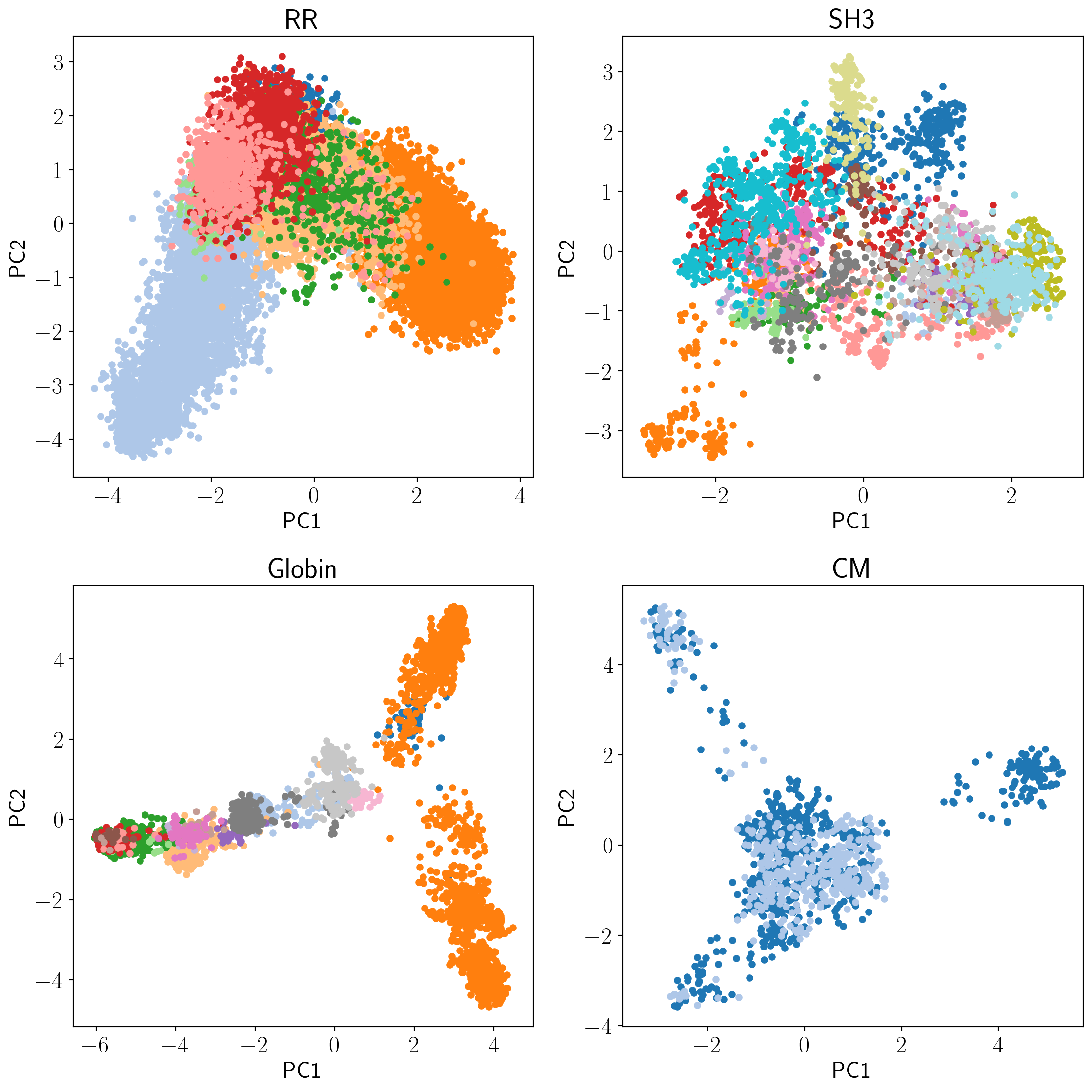}
    \caption{First two projections of the PCA for the entire family of the datasets we considered. The shown PCAs may differ from the ones used for the analysis because the \texttt{cobalt} procedure for constructing the train/test splits leaves out several sequences. For simplicity, we omit the legends with the color to label mapping.}
    \label{fig:PCAs}
\end{figure*}

\begin{figure*}[t]
    \centering
    \includegraphics[width=0.5\linewidth]{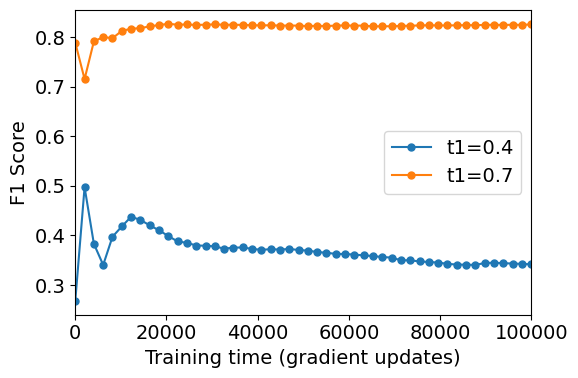}
    \caption{Test set F1 score on the encoding-RBM embedding at different moments during the training. For this figure, we considered the PF00072 dataset with train/test cobalt parameters \texttt{t1}$\in \{0.4, 0.7\}$, \texttt{t2}$=0.7$ and 100 training sequences.
    }
    \label{fig:rbm scores vs time}
\end{figure*}

\begin{figure*}[ht]
    \centering
    \includegraphics[width=\linewidth]{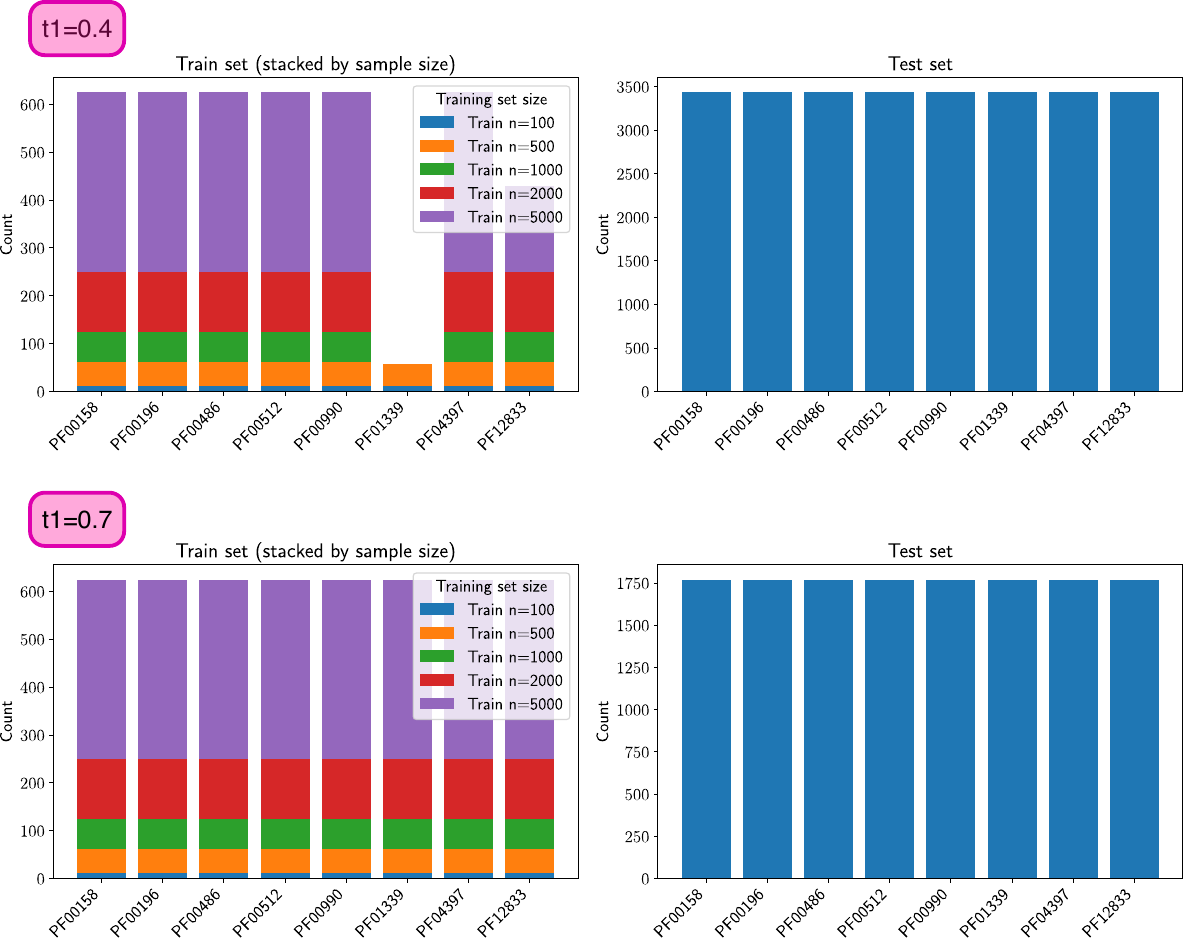}
    \caption{Class abundance for training and test set of RR for different training set sizes and one specific train/test split. The number of training sequences indicated in the legend refers to the target one, but the actual training set size can be smaller than this number.}
    \label{fig:class abundance RR}
\end{figure*}

\begin{figure*}[ht]
    \centering
    \includegraphics[width=\linewidth]{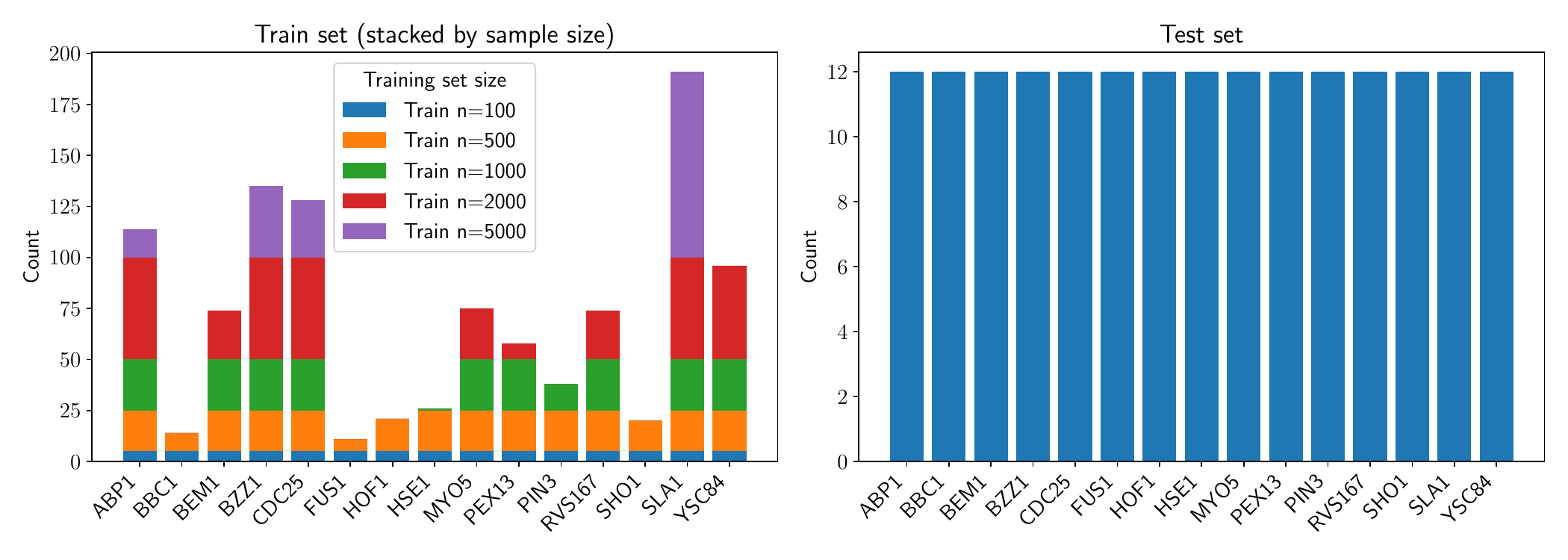}
    \caption{Class abundance for training and test set of SH3 for different training set sizes and one specific train/test split. The number of training sequences indicated in the legend refers to the target one, but the actual training set size can be smaller than this number.}
    \label{fig:class abundance SH3}
\end{figure*}

\begin{figure*}[ht]
    \centering
    \includegraphics[width=\linewidth]{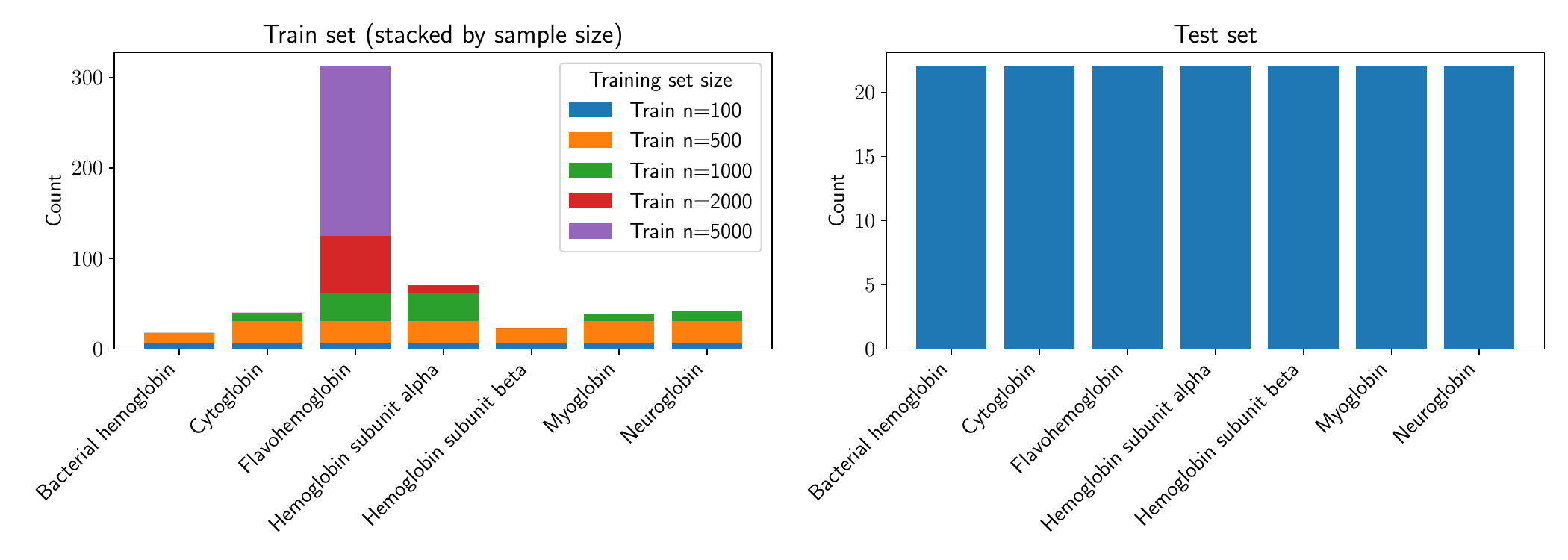}
    \caption{Class abundance for training and test set of the Globin family for different training set sizes and one specific train/test split. The number of training sequences indicated in the legend refers to the target one, but the actual training set size can be smaller than this number.}
    \label{fig:class abundance PF00042}
\end{figure*}

\begin{figure*}[ht]
    \centering
    \includegraphics[width=\linewidth]{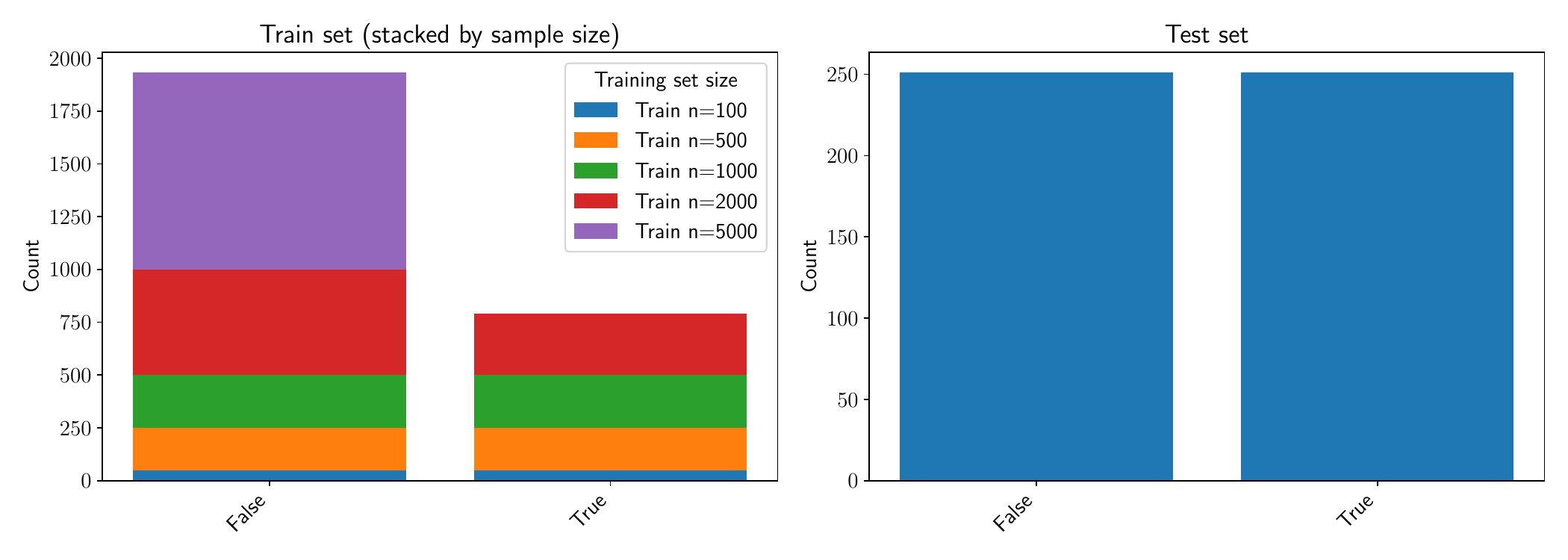}
    \caption{Class abundance for training and test set of CM for different training set sizes and one specific train/test split. The number of training sequences indicated in the legend refers to the target one, but the actual training set size can be smaller than this number.}
    \label{fig:class abundance CM}
\end{figure*}

\begin{figure*}[ht]
    \centering
    \includegraphics[width=\linewidth]{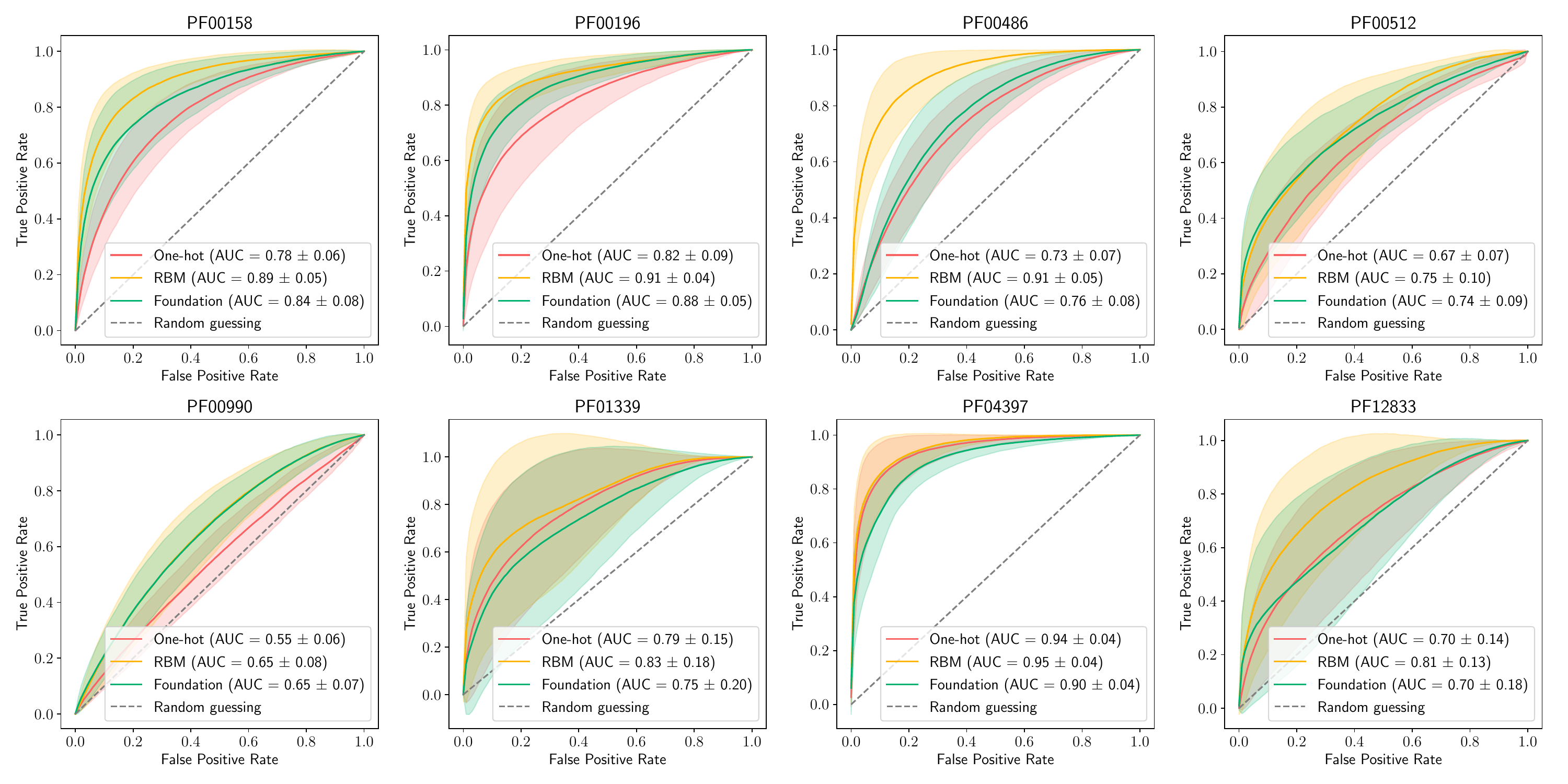}
    \caption{Label-specific ROC curves for RR with \texttt{t1}=0.4 and 100 training sequences. The solid line represents the mean over 10 independent train/test splits, while the shaded area contains one standard deviation.}
    \label{fig:ROC labels PF00072}
\end{figure*}

\begin{figure*}[ht]
    \centering
    \includegraphics[width=\linewidth]{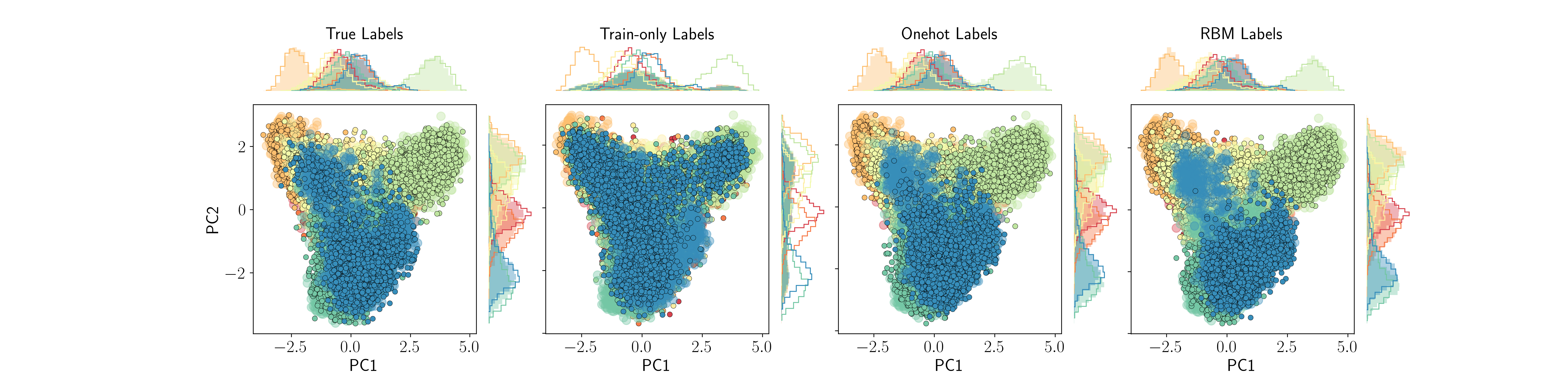}
    \caption{First two PCA projections of natural (large shaded dots and solid-line histograms) and generated (small contoured dots and shaded histograms) data for RBMs informed with different types of label. Note that all data are projected on the first two principal components of the natural MSA, to assure compatibility across datasets. The histograms of natural data, partitioned by true labels, are shown by solid lines, while the solid-filled histograms display the generated data, partitioned by the conditioning label. For this plot we used 100 training sequences and \texttt{t1}=0.7.}
    \label{fig:conditional generation RR t1=0.7}
\end{figure*}

\begin{figure*}[ht]
    \centering
    \includegraphics[width=\linewidth]{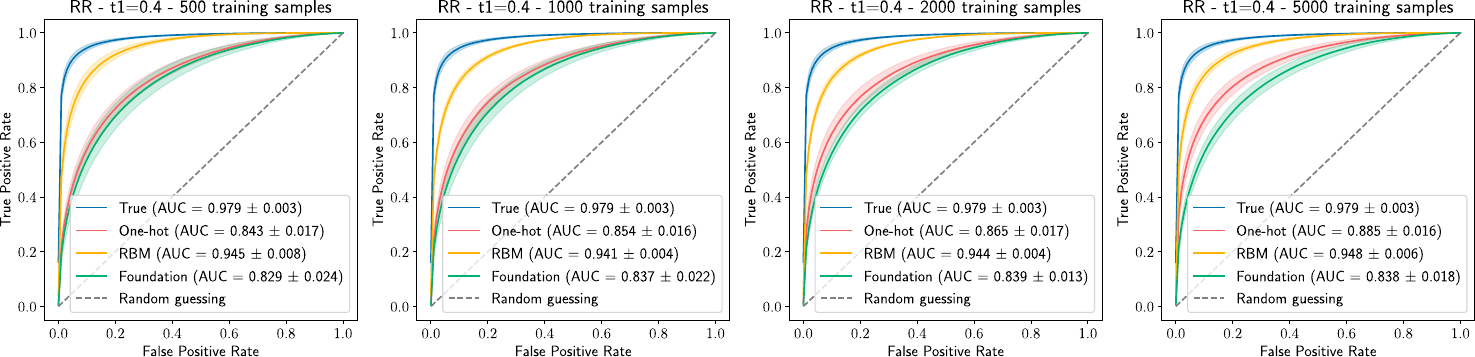}
    \caption{Average ROC curves quantifying the agreement between the label prescription and the label inferred on the generated data by the predictor that informed the RBM for 500, 1000, 2000 and 5000 training samples.}
    \label{fig:self-consistency roc collage}
\end{figure*}

\begin{figure*}[ht]
    \centering
    \includegraphics[width=\linewidth]{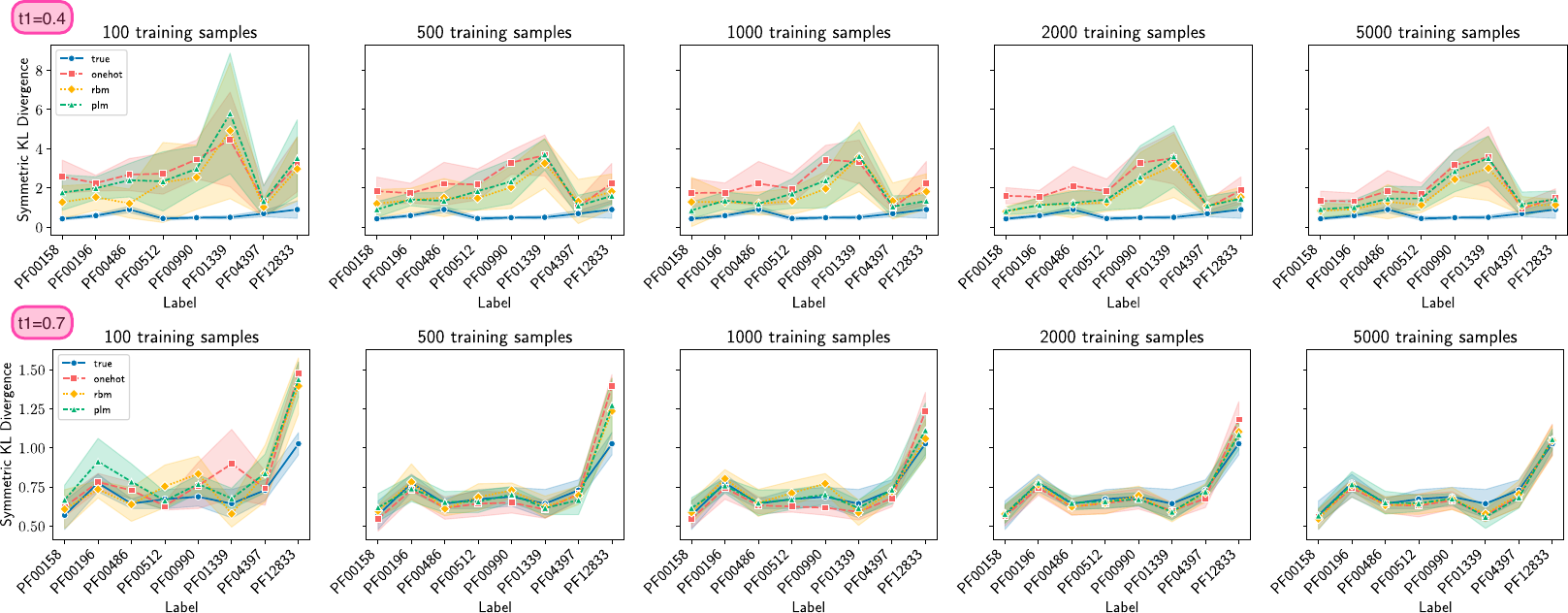}
    \caption{Mean and standard deviation of label-specific symmetric KL divergence between the 2D PCA histograms of natural and conditionally generated sequences for 10 independent train/test splits for the RR dataset. The first row corresponds to \texttt{t1}=0.4 while the second row corresponds to \texttt{t1}=0.7. From left to right, 100, 500, 1000, 2000, and 5000 training sequences have been used for the entire pipeline.}
    \label{fig:Dkl collage}
\end{figure*}

\end{document}